\documentclass[showpacs,preprintnumbers,amsmath,amssymb]{revtex4}

\usepackage{graphicx}
\usepackage{dcolumn}
\usepackage{bm}

\begin{document}

\title{3D-vortex  labyrinths in the near field of solid-state microchip laser.}

\author{A.Yu.Okulov}
 \email{okulov@sci.lebedev.ru}
 \homepage{http://sites.lebedev.ru/okulov}
\affiliation{
P.N.Lebedev Physical Institute of Russian Academy of Sciences \\
Leninsky  prospect  53, 119991 Moscow, Russia }

\date{\ January 15, 2007}

\begin{abstract}

The usage of vortex labyrinths fields and Talbot lattices  
as optical dipole traps for neutral atoms is considered for the wavelength
of trapping radiation in the range 0.98 $\div$ 2.79 $\mu m$. 
The square vortex lattices generated in 
high Fresnel number solid-state microchip lasers
are studied as a possible realization. The distribution of light field is obtained via 
dynamical model based on 
Maxwell-Bloch equations for class-B laser, discrete Fox-Lee map with 
relaxation of inversion and static model 
based on superposition of copropagating  
Gaussian beams. The spatial patterns observed experimentally 
and obtained
numerically are interpreted as nonlinear superposition
of vortices with helicoidal phase dislocations. 
Such patterns are approximated analytically by a sum 
of array of vortex lines and additional parabolic subtrap.
The separable optical trapping potential is proposed. 
The factorization of macroscopic wavefunction have led to 
analytical solution of Gross-Pitaevski equation 
for ensemble of quantum particles trapped in vortex 
labyrinth formed by spatially - periodic 
array of Laguerre-Gaussian beams.

\end{abstract}


\maketitle


\section{\label{sec:level1}Introduction}
		
		Optical dipole traps for neutral atoms \cite{Ovchinnikov:2000} 
are a subject of considerable 
interest.  A certain amount of proposals 
of increasing complexity of trapping EM-field appeared recently. Apart from a relatively simple geometrical 
patterns like standing plane-wave gratings\cite{Pitaevskii:2006}, evanescent  wave 
mirrors\cite{Letokhov:1988}, toroidal traps which utilize the intensity distribution of
Laguerre-Gaussian (LG) beams in beam waist \cite{Wright:2001} a several proposals were made which utilize
the arrays of Gaussian beams, both phase-locked \cite{Okulov:2003,Ovchinnikov:2006} and 
unlocked ones \cite{Ertmer:2002}. The interference inherent to phase-locking
provides multiply  connected configurations of intensity distribution, phase gradients and
electromagnetic (EM) momentum density \cite{Allen:1994,Padgett:2003}. The winding
EM momentum distribution is the cause of the 
angular momentum  transfer to macroscopic bodies (e.g. dielectric ball) \cite{Dunlop} 
and to trapped BEC as well \cite{Abraham:2002}.

	The EM field configuration under consideration is based upon properties of
self-imaging optical fields\cite{Okulov:1990,Courtial:2006}. The difference between phase-locked array 
of zero-order Gaussian beams and  phase-locked array of optical vortices (OV) obtained experimentally
in the near field of solid-state 
microchip laser \cite{Chen:2001} is that the latter consists of array of parallel vortex lines with opposite 
circulations and topological charges $\ell_{EM} $ (TC)  (fig.  \ref{fig.1}) \cite{Okulov:2003}.
In contrast to earlier proposals where individual loading and addressing 
of trapping sites was considered \cite{Ertmer:2002}  this OV array have sophisticated configuration 
of intensity \cite{Okulov:2003} and EM momentum density of trapping field.  
Due to this configuration
the BEC trapped in EM field  might have 
macroscopic wavefunction of complex form composed of array of superfluid vortices (SFV). 
The  cause of SFV formation is light-induced torque experienced by isolated 
resonant atom which interacts with Laguerre-Gaussian beam having phase singularity. 
It was show by Allen et al.\cite{Allen:1994} the value of torque 
in saturation limit is $T= \hbar{\:}  \ell_{EM} {\:} \Gamma $. The  origin of  torque  is due to nonzero 
azymuthal EM momentum component. The azimuthal Doppler shift corresponding to such motion 
had been onserved\cite{Padgett:2003}. This torque might have an appreciable value 
even in nonresonant case, although it is significantly reduced by the multiple of $ ({\Delta} /{\Gamma})^2 $,
where $\Delta $ - is detuning, $ \Gamma $ - is linewidth\cite{Allen:1994}. As a result an azimuthal 
component of EM momentum is transferred 
to atom in such a way that it will move around phase singularity - 
the direction of rotation is fixed by TC of trapping beam. This is the cause of 
circular motion of BEC confined in isolated toroidal trap. 

In current paper the maximally transparent and simple analytical 
description presented which describes the transfer of angular momentum 
from optical vortex array (OVA)  to trapped BEC.  As a first step in $section {\:} 2$  the distribution 
of light intensity and phase in the near field of solid-state microchip
laser are computed.
The configuration of intensity and phase of EM-field of OVA  has complicated, multinode structure.  In $sections {\:} 3 {\:}  $ 
and  $section  {\:}  4 $ 
the spatially periodic arrays of Gaussian and Laguerre-Gaussian beams of the first order are compared. Next in $section {\:} 5 $ 
in order  to 
get analytic solution for macroscopic wavefunction $\Psi$ trapped by optical vortex array,
the special optical  pancake-like potential  (fig. \ref{fig.1}) will be constructed. In this trapping potential
the separation of variables 
in Gross-Pitaevsky equation\cite{Pitaevskii:2006} becomes possible. Next the procedure applied previously to elongated $sech^2$-
profile optical trap \cite{Manko:2004} will be used. Due to special ajustment of potential it is possible to 
capture an analytic approximation for $\Psi$ in the form of superposition of elementary equispaced vortices.
The topological charges and angular momenta of adjaicent vortices are counter-directed in
contrast to  "rotating  bucket" trap where angular momenta of SFV in bucket are 
co-directed \cite{Ketterle:2002,Feynman:1972}.  
The numerical modeling via split-step FFT technique will be performed to check this approximate formula.
The obtained spatial distribution of $\Psi$ modulus and argument will be used for analysis of 
classical field of velocities of a trapped atom. The complex form  
of constructed macroscopic wavefunction $\Psi$ trapped by optical vortex array might be interesting
 from the point view of diminishing
the decoherence induced by environment in topological quantum computing\cite{Kitaev:2001}. 

\section{\label{sec:level1}Square vortex lattices}

			Recent advances in controlling the dynamics of solid-state microchip lasers  \cite{Chen:2001} offer the 
possibility of reliable control of spatiotemporal optical patterns. Compared to semiconductor lasers
the host medium composed of dielectric crystal doped by neodimium $Nd^{+3}$ or other 
rare earth ions ($Er^{+3},Tm^{+3},Ho^{+3},Yb^{+3}$) have smaller gain and smaller  
self-phase modulation. Because of smaller nonresonant losses the heating of host crystal by 
radiation is not dramatic. The changes of the geometry 
and birefrigence of host crystals, curvature and reflectivity of output couplers and spatial distribution of optical pumping 
give efficient control over mode structure.
The square vortex lattices (SVL) observed in quasi plane parallel cavity\cite{Chen:2001} are of 
special  interest. These lattices demonstrate  high degree of spatial coherence
 : the relaxation oscillations of class-B high Fresnel number solid-state microchip 
laser \cite{Chen:2001} with Fresnel number in the range $N_f \approx 100 \div 1000$ are characterized by 
single peak at frequency about  $({\sqrt {T_1 {\tau_{c}}}})^{-1}$ ( $ 0.3 \div 1.2 {\:} MHz$ ).This is the firm evidence of single-longitudinal and single transverse mode behaviour.
The former theoretical analysis was based on parabolic equation, resulted from adiabatic elimination of polarization from
standard set of Maxwell-Bloch equations \cite{Suchkov:1965}:
\begin{eqnarray}
\label{she_staliunas}
\ {\frac {\partial E(\vec{r},t )}{\partial {\:} t}} + { \frac {\ E(\vec{r},t )}
{\tau_c}}+{ \frac {\ i {\:}{\:}c }{2 k }} {\Delta _{\bot}E(\vec{r},t ) } =
\ {\frac {\sigma {c N_0 L_a} E(\vec{r},t )(1 + i \delta \omega {\:}T_{2}) }{{2 L_r}\ (1 + \sigma T_{1}{\:}c{\:} \epsilon_0   |E|^2 / { \hbar{\omega }})}} {\:}
\end{eqnarray}

where ${\tau_c}$ - is photon lifetime in cavity, $ k $=${2 \pi}{/{\lambda}}$ - wavenumber, 
${\sigma }$ - stimulated emission crosssection, ${\delta \omega }$ - detuning of lasing frequency from center of gain line,  
$ N_0$ - density of inverted resonant atoms per unit volume, $ {T_1} $ - inversion lifetime(longitudinal relaxation lifetime), 
$L_a$ - thickness of active medium, $L_r$ - length of resonator, $c$ - speed of light,$ {\:}{\:}\epsilon_0 $ - dielectric constant,
${\Delta _{\bot}}$=${{\nabla_{\bot}}^2}$ \cite{Siegman:1986}.
When finite gain linewidth $ {T_2}^{-1} $ is taken into account in the framework of Swift-Hohenberg equation:

\begin{eqnarray}
\label{she_staliunas}
\ {\frac {\partial E(\vec{r},t )}{\partial {\:} t}} + { \frac {\ E(\vec{r},t )}{(\tau_c+{T_2})}}+{ \frac {\ i {\:} {\tau_c} {\:}c }{2 k (\tau_c+{T_2})}} {\Delta _{\bot}E(\vec{r},t ) } 
&-& \nonumber \\
{\frac {{{T_2}^2}}{{\tau_c}({\tau_c}+{T_2})^2}} ({ \frac {\tau_c{\:} c }{k}} {\Delta_{\bot} + {\delta {\omega {\:}T_{2}}}})^2{\:}E(\vec{r},t )=
\ {\frac {\sigma {c N_0 L_a} E(\vec{r},t )(1 + i \delta \omega {\:}T_{2}) }{{2 L_r}\ (1 + \sigma T_{1}{\:}c{\:} \epsilon_0   |E|^2 / { \hbar{\omega }})}} {\:}
\end{eqnarray},
the square vortex lattices were obtained
numerically\cite{Staliunas:1993}.
The additional term with the square of transverse laplacian is responsible for transverse mode selection 
due to finite gain linewidth.An alternative model with discrete time step equal to $\tau=2 L_r/c$  (time of bouncing of radiation between mirrors)
utilizes the standard rate equations of class-B laser written  at $n$-th step for electric field\cite{Okulov:2003,Hollinger:1990}:
\begin{equation}
\label{nonlin}
\ E_{n+1}(\vec{r})=f(E_{n}(\vec{r})) = \frac{\sigma L_a N_{n}(\vec{r}) E_{n}(\vec{r}) (1+i{\delta \omega}T_2) }{2}+E_{n}(\vec{r}),
\end{equation}
inversion:
\begin{equation}
\label{kernelfabry2}
\ N_{n+1}(\vec{r}) = N_{n}(\vec{r})  +[{\frac{N_{0}(\vec{r})-N_{n}(\vec{r})) }{T_1}}-{\sigma} N_{n}(\vec{r}) {\:} 
c{\:}   \epsilon_0  {\:}  |{E_{n}}|^2 / { \hbar {\omega }} ]{\frac{2 L_{r}}{c}},
\end{equation}

and nonlocal integral mapping evaluating the field via fast Fourrier transform
at each timestep:
\begin{equation}
\label{nonlmap3}
\ E_{n+1}(\vec{r} )= \int^\infty_{-\infty}\int^\infty_{-\infty}\ K(\vec{r}-\vec{r'}) f(E_{n}(\vec{r'})) d^2 \vec{r'},
\end{equation}
The kernel  $ {K}$ for  the  nearly  plane-parallel Fabry-Perot cavity of microchip laser with 
transverse filtering via aperture $D(\vec{r'})$ has the form \cite{Okulov:1990} :
\begin{equation}
\label{kernelfabry3}
\ K(\vec{r}-\vec{r'} )= \frac{ikD(\vec{r'})}{2 \pi {{L_r}} }   {\:}  exp{\:}  [ik(\vec{r}-\vec{r'})^2 / 2{{L_r}}].
\end{equation}

The following parameters were choosen in numerical simulation: $ {T_1}   = 2 { \cdot} 10^{-4}sec $,   
$L_r =  1 mm$, for $Nd^{+3}$-doped crystals ${\sigma }  =  {(1.2-0.6)}{ \cdot} 10^{-20} cm^2$,
$ N_0  =   10^{16} cm^{-3}$, ${{ \delta \omega}{T_2}   =   {0.1}  }$.
In numerical evaluation of eq. (\ref{nonlmap3}) via split-step FFT method \cite{Okulov:1993}
the mesh size in $X,Y$ plane   was $512 \cdot 512$ points.
The "guard bands ratio" \cite{Siegman:1975} was set equal to $8$. The main part of the field
$ \Psi $ was located inside central part of a mesh of  $64 \cdot 64$ size - the "image area". The tolerance 
of the energy spillover was kept within $\epsilon_1 = 0.001$.
The  "windowing" in 
wavenumber space after FFT at each timestep was performed by usage of "Fermi-Dirac" smoothed 
step function \cite{Okulov:1993}.
The dissipation inherent to "windowing" in split-step FFT 
method was compensated by nonlinear gain. 
The initial conditions for field $E_{n}$ were taken as multimode random field \cite{Okulov:1993}.

Just near lasing threshold the radiation mode 
has a distribution of intensity as rectangular grating of bright and dark spots : the latter are vortex cores.
The nonlocal integral mapping proved to be succesful in computation of the near field distribution as well. 
Quite unexpectably in the most runs the parallel vortex lines were
obtained  (fig. \ref{fig.1}) \cite{Okulov:2003} rather than periodic array of bright and dark spots, 
typical to Talbot phenomenon\cite{Okulov:1990}. The origin of parallel vortex lines  
is  interpreted as nonlinear superposition
of vortices with helicoidal phase dislocations. 
In the next  $section {\:} 3{\:}  $ and  $section  {\:} 4 $ 
 the possibility of approximation of 
this nonlinear optical vortex lattice by linear equivalent will be outlined.

\section{\label{sec:level1}Talbot lattices}

Consider the phase-locked rectangular lattice  of zero-order 
Gaussian beams  located at sites $\vec{r}_{jx,jy}$ \cite{Okulov:1990} separated by period $p$,
where $jx,jy$ - are dicrete indices corresponding $x$ and $y$ coordinate of a given site.
Let us assume for simplicity that polarization is linear, thus spin 
of light is zero.
At $z=0$ plane the electric field $E$ is given by expression:
\begin{equation}
\label{tlb3}
\ E(\vec{r},0 )= E_0 \sum_{jx,jy} {\:} exp{\:}[ - {| \vec{r}-\vec{r}_{jx,jy} | }^2/(2 d^2)]
\end{equation}

After paraxial propagation of distance  $z$  the electric field $E(\vec{r},z )$ is transformed into:
\begin{equation}
\label{tlb4}
\ E(\vec{r},z )= E_0 {\:} {\frac {{{i}}{\:} exp{\:}[{\:} {{i}} k z]}{ (1  -  {{{i}} z/k d^2}) }} 
\sum_{jx,jy} {\:} exp{\:}[- \frac {{| \vec{r}-\vec{r}_{jx,jy} | }^2}{2 d^2(1 -  {{i}} z/(k d^2))}]
\end{equation}

The constructive interference between ajaicent beams produces periodic interference 
patterns in diffrenent  z-spaced planes. The initial periodical pattern  
is reproduced at so-called Talbot-distances $z_t =2 p^2/{\lambda}$. In the intermediate planes 
$z_t/4m$ the coarser lattices with periods p/m are produced\cite{Okulov:1993,Okulov:1994}. 
As a result a 3D lattice of bright spots is formed (fig. \ref{fig.2}).

Each spot could serve as potential well for neutral atoms 
\cite{Ovchinnikov:2000,Pitaevskii:2006,Okulov:2003,Ovchinnikov:2006},
because intensity gradient will attract or repulse the atomic dipoles depending the sign
of detuning of radiation frequency from resonance. 
At low frequencies(red-detuning) atomic dipole oscillates in-phase with trapping field 
and tries to align parallel to electric field.Thus potential energy of dipole $U=-{\vec p}{\cdot}{\:}{{\vec E}(\vec{r} ) }/2$ is lower 
in local maxima of intensity and atoms are collected at bright spots.
On the other hand at frequencies above the resonance (blue-detuning) atomic dipole 
oscillates out-of-phase
and it has tendency  to align anti-parallel to electric field. In this blue-detuned case 
the potential energy of dipole is higher 
in local maxima of intensity and atoms are repelled into dark regions.

Such geometry of trapping in integer and fractional Talbot planes based on superposition of co-propagating  
 zeroth-order Gaussian beams was considered earlier\cite{Okulov:2003} including possibility of manipulation
of optical lattice geometry via mutual polarization \cite{Ovchinnikov:2006} of constituting beams.

\section{\label{sec:level1}Artificial vortex labyrinth}
Consider now the periodic array of Laguerre-Gaussian vortex beams with helicoidal 
phase dislocations (fig. \ref{fig.3}):
\begin{eqnarray}
\label{hold_vort}
\ E(\vec{r},z=0 ) \approx  E_0 \sum_{jx,jy}{(| \vec{r}-\vec{r}_{jx,jy} |)}{ exp{{\:}(- {\:} {| \vec{r}-\vec{r}_{jx,jy} |}^{{\:}2}/d^2})}
& &  \nonumber \\ { exp{{\:}(- {\:} {| \vec{r} |}{\:}^2/D^2})}{\:} {exp{\:}[ {\:}{{i}}{\:} \ell_{EM} Arg( \vec{r}-\vec{r}_{jx,jy} ){\:}+ {\:}{{i}}{\:} \pi (jx+jy)}].
\end{eqnarray}

The topological charge $ \ell_{EM}$ is assumed to be unity, the neighbouring beams (components of the sum (\ref{hold_vort}))
are $\pi$-shifted. The apodization function $ { exp{{\:}(- {\:} | \vec{r} |{\:}^2/D^2})}$ is added 
in order to suppress the maximum of interference pattern at the edge of array.
The beams centers are placed in the centers of rectangular 
grid $\vec{r}_{jx,jy}$ of period $p$ whose axes are parallel to $X$,$Y$. The overlapping beams produce interference pattern formed by 
two arrays of bright and dark spots rotated at ${45^{\circ}}$ angle with respect to initial array 
of LG beams. The dark spots are of two kinds: one lattice of spots coincides with lattice
of initial vortices(\ref{hold_vort}), the other one is produced by interference  and it is
shifted at distance $p/{\sqrt{2}}$ along diagonal of initial lattice. The resulting interferogram has apparent
 ${45^{\circ}}$ tilt compared to lattice of initial vortices (fig. \ref{fig.3}).

The topological charges of dark spots (vortices) 
flip from one site to another. The interesting feature of this interference pattern 
is the distribution of angular momentum \cite{Courtial:2006}.   The initial array of 
LG beams carries unit circulation and corresponding angular momentum at each 
site. Such $chessboard-like$ interferogram  (fig.  \ref{fig.3}) contains additional array of vortices with alternative
charges. The net angular momentum tends to be close to zero, because 
at the central part of array each positively directed TC  $ \ell_{EM}$ is compensated by 
four adjaicent negative ones having charge ${\:}- \ell_{EM}$ . The elementary cell of such a lattice consists of two $\pi$-shifted 
initial vortices with co-directed charges $ \ell_{EM}$  located at diagonal and two 
counter-directed  charges  ${\:}- \ell_{EM}$ ($\pi$-shifted too) placed at the other diagonal of cell.
The period $p$ of initial pattern in numerical recipes \cite{Okulov:2003} was taken as $30 \mu m$ while
width of each initial beam was set ${w_{LG}}=22  {\mu m} $ in order to provide significant 
mutual overlapping of vortices . 

The longitudinal distribution of intensity of optical vortex array is composed of periodically spaced
(with period $p$) hollow tubes - vortex cores(fig. \ref{fig.3}). Apart from Talbot gratings which are 
reproduced by diffraction at $z_{t} ={2 p^2}/ \lambda$ planes with corresponding period division 
in between planes the vortex array under consideration keeps its shape within Rayleigh range,
i.e. at distances $z<{D^2}/ {\lambda}$. Within this distance the diameter of cores is kept 
constant along $z$ by virtue of interference of adjaicent
vortices, whose helical wavefronts are perfectly matched within elementary cell (fig. \ref{fig.4}).

Consider now the interaction of individual atoms with single optical beam carrying topological charge. 
The gradient force will attract the "red"-detuned atomic dipole to the intensity  maximum of 
an isolated first order Laguerre-Gaussian beam  - "doughnut" beam (fig. \ref{fig.4}), i.e. to the 
ring-shaped area around phase singularity.
As a result the cloud  of cold atoms will be accumulated near the maximum of intensity or  "doughnut" 
as follows from  variational 
solution of Gross-Pitaevsky equation (GPE) \cite{Pitaevskii:2006}:
\begin{eqnarray}
\label{GPE2}
\  {i \hbar}{\:}{\frac {\partial {\Psi}(\vec{r},t )}{\partial t}} =
 -{\frac {\hbar^2}{2 m}} \Delta {\Psi}(\vec{r},t )+V_{ext}(\vec{r}){\Psi}(\vec{r},t )
& +&    \nonumber \\
 + {\frac{4{\pi}{\hbar}{\:}^2{\:} {a}}{m}}
 {\Psi}(\vec{r},t )|{\Psi}(\vec{r},t ) |{\:}^2 ,
\end{eqnarray}

with  trapping potential ${V_{ext}}$ of the form 
\begin{equation}
\label{trapLG}
\ {V_{ext}}(\vec{r} )= {\frac { m{\:} {\omega{_z}}^2 {z^2}}{2 }}-Re[{\alpha(\omega)] {\:} }
{\:}|{E_0} |{\:}^2{\:} [ {r}^2 {\:}exp{\:}(-{r} {\:}^2/{(2{w_{LG}}^2)}) ]
\end{equation}

where $\alpha(\omega) $ is polarizability of atom\cite{Ovchinnikov:2000}, $m$ - mass of particle, ${a}$ - scattering length.

The approximate variational 
solution of GPE for ground state macroscopic wavefunction is 
Laguerre-Gaussian function \cite{Abraham:2002}:

\begin{equation}
\label{LG}
\ {\Psi}(\vec{r},t )=  | \vec{r} | {\:}exp{\:}(- {| \vec{r} |{\:}^2}/{(2 {w_r}^2)}-z^2/{(2{w_z}^2)}+{{i}}{\:}{\ell_{BEC}} {\:} \phi),
\end{equation}
where ${\ell_{BEC}}$ is the topological charge of this vortex $\Psi$ solution.
Hence the  probability distribution $|{\Psi}(\vec{r},t ) |{\:}^2$
of finding atom some  near the point ${\vec{r}=(x,y,z)}$ at the moment $t$ is similar to the intensity 
distribution of trapping field as pointed out in \cite{Abraham:2002}. More information 
could be obtained from the study of phase structure of wavefunction and comparison of topological charges of
trapping field and BEC vortex.
It is easy to realize using Madelung transform
\begin{equation}
\label{MT}
\ {\Psi}(\vec{r},t )= \sqrt { \rho( | \vec{r} | , \phi ,t ) } {\:}exp{\:}({{i}}\theta ( | \vec{r} | , \phi, t ) )
\end{equation}
 that there exists  flow of "probability fluid" with velocity $\vec{v}$ proportional (parallel) to the  
phase gradient lines:
\begin{equation}
\label{velo}
\ v(\vec{r},t )= {\frac { \hbar}{m }} \nabla \theta(\vec{r},t ) 
\end{equation}

Such "flow of wavefunction" occurs around the $z$-axis (beam axis).
The "flow" described by  (\ref{LG}) is potential and conservative 
as it should be for superfluid. This picture is complicated by vorticity of EM-momentum,
inherent to LG-beams \cite{Padgett:2003}.
Strictly speaking the rotation of classical dipole around core is accelerated by nonconservative torque induced by 
azimuthal force circulating around vortex core. For two-level atom the value of torque $T$ is given by following 
expression obtained by Babiker, Power and Allen \cite{Allen:1994}:

\begin{equation}
\label{torque}
\ T = {\hbar{\:}  \ell_{EM}{\:} \Gamma} {\:} [ {\frac {  {I_{L}} }{1+{I_{L}}+{{\Delta}^2}/{{\Gamma}^2} }} ] 
\end{equation}

where ${I_{L}}$ - normalized trapping filed intesity, 
${\Delta} = {\omega - \omega_0} $ - detuning of trapping field  frequency ${\omega }$ 
from resonant frequency ${\omega_0 }$ of dipole. In saturation limit $T$ is simply : $T= \hbar{\:}  \ell_{EM} {\:} \Gamma $. 
In nonresonant case this torque reduces
as ${\Gamma}^2/{\Delta }^{2}$. At trapping frequencies of $Nd^{+3}$ laser ${\Delta =10^{14}}Hz$ and $GaAs$ laser ${\Delta =10^{13}}Hz$
the $D$-line dublett of $^{85}Rb$ with linewidth ${\Gamma = 5{\cdot}10^{{\:}6}}Hz$ \cite{Ovchinnikov:2000}
the torque is reduced by factor depending on laser intensity  ${I_{L}}$ .Thus due to 
trade-off between the saturation and detuning the torque exists both near and at large detuning from the 
resonance. Consequently the loop integral of the azymuthal force
over circular trajectory around core $\oint({{\vec{F_t}} \times { \vec{r}}})\cdot d { \vec{l}}$
is nonzero (fig.  \ref{fig.4}).

In classical picture because of this torque  the
dipole  placed in the doughnut beam will move "upwards"
the helicoidal phase staircase, i.e. it will rotate around LG-beam axis. The direction of 
rotation is determined by  topological charge $ \ell_{EM}$ of the trapping beam  (fig. \ref{fig.4}). Qualitatively classical dipole 
is pushed by azimutally periodic electric field - it happens because the phase of electric field oscillations at 
each point in circle around the center of beam is shifted with respect to neighbouring points. 
Thus classical dipole placed at maximum of intensity of GL-beam will feel the 
"plane-wave-like" nonresonant pressure of light field. The associated azimuthal Doppler shift 
of the moving atom was observed experimentally recently \cite{Dholakia:2002}.
Outside the resonance the origin of torque $T$ is
interpreted  in similar way: the Pointing vector has component $S(\vec{r})_{t}$ tangent to helix and local
flux of electromagnetic momentum pushes dipole along phase gradient, i.e. in azimuthal direction.
The local density of EM-momentum $\vec{g} =\vec{ S}/c^2$ is proportional to components 
of Pointing vector\cite{Padgett:2003}:
\begin{eqnarray}
\label{em_momentum}
\   S(\vec{r})_{t}= {\frac {{\epsilon_0}  {\:}\omega  {\:} \ell_{EM}{\:} {c^2}}{r}{  }}  { |{E}(\vec{r} ) |{\:}^2 } 
& &    \nonumber \\ S(\vec{r})_z= {\epsilon_0} {\:}c {\:}{ |{E}(\vec{r} ) |{\:}^2 },
\end{eqnarray}
where $ S(\vec{r})_{t}$ - is tangential component of Pointing vector,  $ S(\vec{r})_{z}$ - 
is axial one, $|(\vec{r} ) |$  - is the distance from optical vortex core, $\omega$ - is frequency of 
trapping field. The existence of tangential component of Pointing vector became visible
when macroscopic dielectric 
ball with radius larger than LG-beam core and comparable 
to "doughnut" radius is placed in center of LG-beam.  The rotation of particle is induced 
via such torque and corresponding transfer of the angular momentum\cite{Dunlop}.

When loaded in optical vortex lattice  (fig.  \ref{fig.5})the atomic dipole will move 
around the adjaicent vortex core with acceleration. The radius of
rotation will increase until the dipole will approach the separatrice of the velocity 
fields.Next after certain amount of rotations around the vortex core 
it could jump to another vortex using bright areas between vortices as bridge (fig. \ref{fig.4}).
This qualitative picture is complicated 
by azimuthally inhomogeneous distribution of intensity around each core.	

Our aim now is to show that just described classical motion have quantum mechanical counterpart. The 
transfer of angular momentum results in the specific form of macroscopic wavefunction $\Psi$
maintaining coherence all over trapping array:
\begin{eqnarray}
\label{psi_vortex_array}
\ {\Psi( \vec{r})}    \approx
 \sum_{jx,jy} {(| \vec{r}-\vec{r}_{jx,jy} |)}{ exp{{\:}(- {\:} | \vec{r}-\vec{r}_{jx,jy} |^{{\:}2}/d^2})}
& &    \nonumber \\ { exp{{\:}(- {\:} | \vec{r} |{\:}^2/D^2})}{\:} {exp{\:}[ {\:}{{i}}{\:} \ell _{BEC}  Arg( \vec{r}-\vec{r}_{jx,jy} ){\:}+ {\:}{{i}}{\:} \pi (jx+jy)}].
\end{eqnarray}

Next section presents the method of solution of Gross-Pitaevsky
equation with vortex trapping field $ E(\vec{r})$  in the form  (\ref{hold_vort}).

\section{\label{sec:level1}Separable vortex array potential for BEC}
	In order to get closed form solution for macroscopic $\Psi$ it is worth to mention 
that azimuthal accelerating force has very small value, falling as ${\Delta}^{-2}$ under 
detuning from resonance\cite{Allen:1994}.
Next let us introduce optical potential $V_{ext}$ as a square modulus 
of trapping field ${E}(\vec{r_{\bot}},z )$\cite {Ovchinnikov:2000} . The torque $T$ will be taken into 
account as a "selection" rule 
for choosing distribution of topological charges $l_{BEC}$ in resulting solution.

	It was shown recently \cite{Manko:2004} that Gross-Pitaevsky equation\cite{Pitaevskii:2006} in 3D-configuration: 
\begin{eqnarray}
\label{GPE3}
\ {{{i}} \hbar}{\:}{\frac {\partial {\Psi}(\vec{r},t )}{\partial t}} =
 -{\frac {\hbar^2}{2 m}} \Delta {\Psi}(\vec{r},t )+V_{ext}(\vec{r}){\Psi}(\vec{r},t )
& +&    \nonumber \\
 + {\frac{4{\pi}{\hbar}{\:}^2{\:}{a({\vec{B}})}}{m}}
 {\Psi}(\vec{r},t )|{\Psi}(\vec{r},t ) |{\:}^2 ,
\end{eqnarray}
admits the application of standard method of separation of variables 
widely used for solution of $linear$ partial differential equation, e.g. in quantum mechanics. 
The separation of variables means that wavefunction is factorized:

\begin{equation}
\label{PSISEP}
\ {\Psi}(\vec{r},t )={\Psi_{\bot}}(\vec{r}_{\bot},t ){\Psi_{||}}(z,t )
\end{equation}

provided the Hamiltonian is the sum of two components. First component depends on a longitudinal variable $z$
and a second component depends upon  transverse variables  $\vec{r}_{\bot}$.

Following to \cite{Manko:2004} in order to separate  variables and factorize the wavefunction 
let us choose trapping potential in the following form, as a sum of components depending on longitudinal coordinate
$z$ and transverse coordinates $\vec{r}_{\bot}$ separately:

\begin{eqnarray}
\label{TRAPOP}
\ { V_{ext}(\vec{r_{\bot}},z)}=V_z +V_{_{\bot}} = {\frac { m{\:} {\omega{_z}}^2 {z^2}}{2 }}-Re[{\alpha(\omega)] {\:}
} |{E}(\vec{r}_{\bot} ) |{\:}^2
& +&    \nonumber \\
+ {\frac { m{\:} {\omega{_{\bot}}}^2 { | (\vec{r}_{\bot} ) |{\:}^2  } }{2} }
\end{eqnarray}

where  ${\alpha(\omega)  }$ is atomic polarizability\cite{Ovchinnikov:2000} : 

\begin{equation}
\label{polar}
\alpha(\omega) = 6 \pi \epsilon_0 {\:} c^3  {\frac {\Gamma / {\:}{\omega_0}^2 {\:} } {(\omega_0^2-\omega^2-{{i}}(\omega^3 / {\omega_0}^2)\Gamma)}}
\end{equation}
The ${\alpha(\omega)  }$ is assumed to be real due to large negative ("red") detuning from 
atomic resonance at frequency ${\omega_0=2.4{\cdot}10^{15}} $Hz. The imaginary part of denominator under detuning 
${\Delta} = {\omega - \omega_0} $ for $^{85}Rb$ atoms
trapped by EM-field at $\lambda = 1.06  \div  0.808  {\:} \mu m$ is $(3{\div}4.7){\cdot}10^8$ times smaller than real part, so the 
permittivity of atom ${\alpha(\omega)  }$ is real with good accuracy \cite{Ovchinnikov:2000}. 
The trapping field $E(\vec{r_{\bot}})$ is assumed to be periodic function of transverse variables $\vec{r_{\bot}}=(x,y)$,
composed of LG beams placed at the nodes ($jx,jy$) of rectangular grid of period $p{\:}{\:}$ (\ref{hold_vort}) .

\begin{figure}
\center{\includegraphics[width=.6\linewidth] {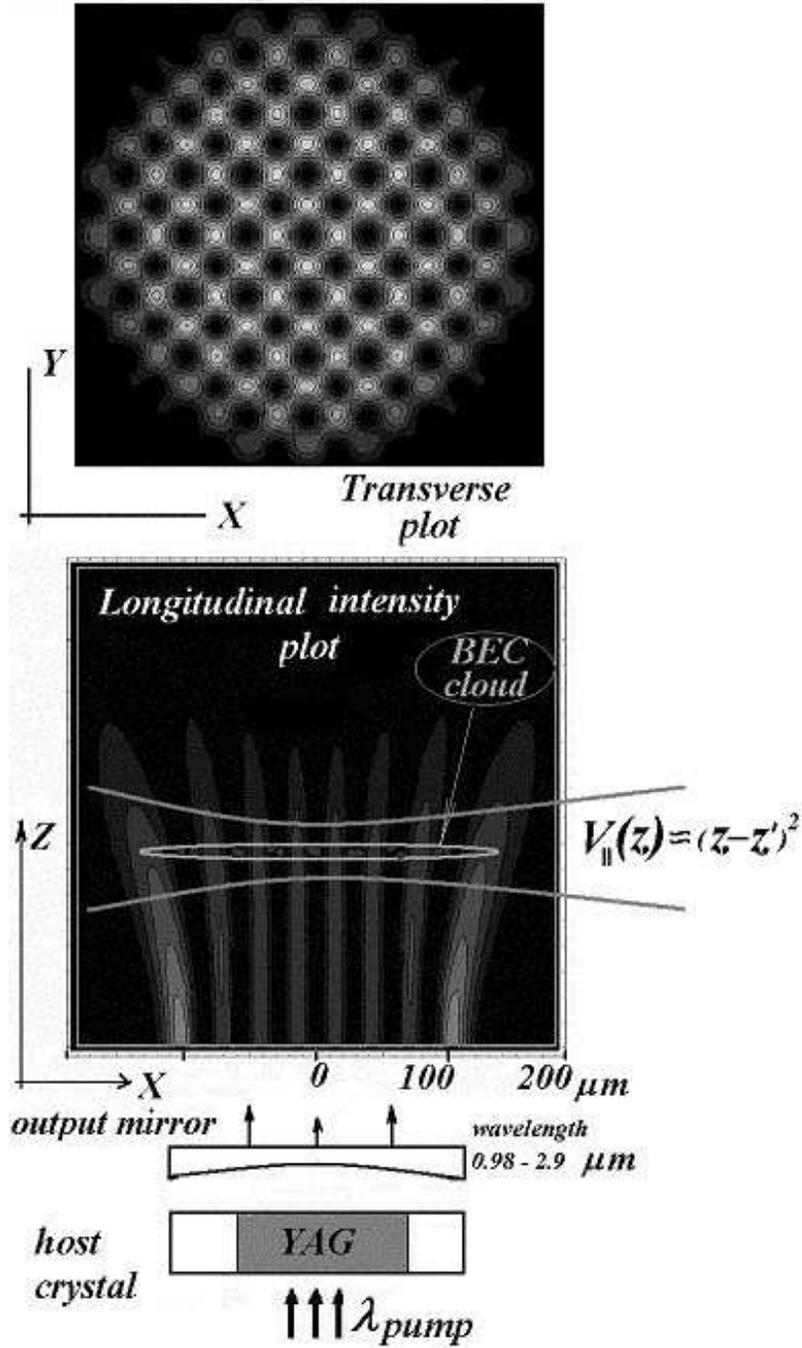}}

\caption{Conceptual view of the near field optical trap. Upper plot:
 Transverse (in XOY plane) distribution of intensity in the near field of
solid-state microchip laser {\cite{Okulov:2003}}. 
Middle: The longitudinal scale extends to 6 Talbot lengths. 
Z - axis is directed along optical axis of microchip laser 
resonator(bottom) . Additional tightly confined parabolic well
keeping BEC cloud localized in $z$ - direction is depicted 
via potential $V(z) \approx (z-z^{\prime})^2$. Such potential is assumed to be 
superimposed by the other microchip laser beam with cylindrical focusing lens at slightly different
wavelength 
of radiation from the range ${0.98 \div 2.79 {\mu} m}$, in order to avoid interference.
 Fresnel number has value of $N_f \approx {N_{vortices}} ^2    =  64$.}
\label{fig.1}
\end{figure}

\begin{figure}

\center{\includegraphics[width=.6\linewidth] {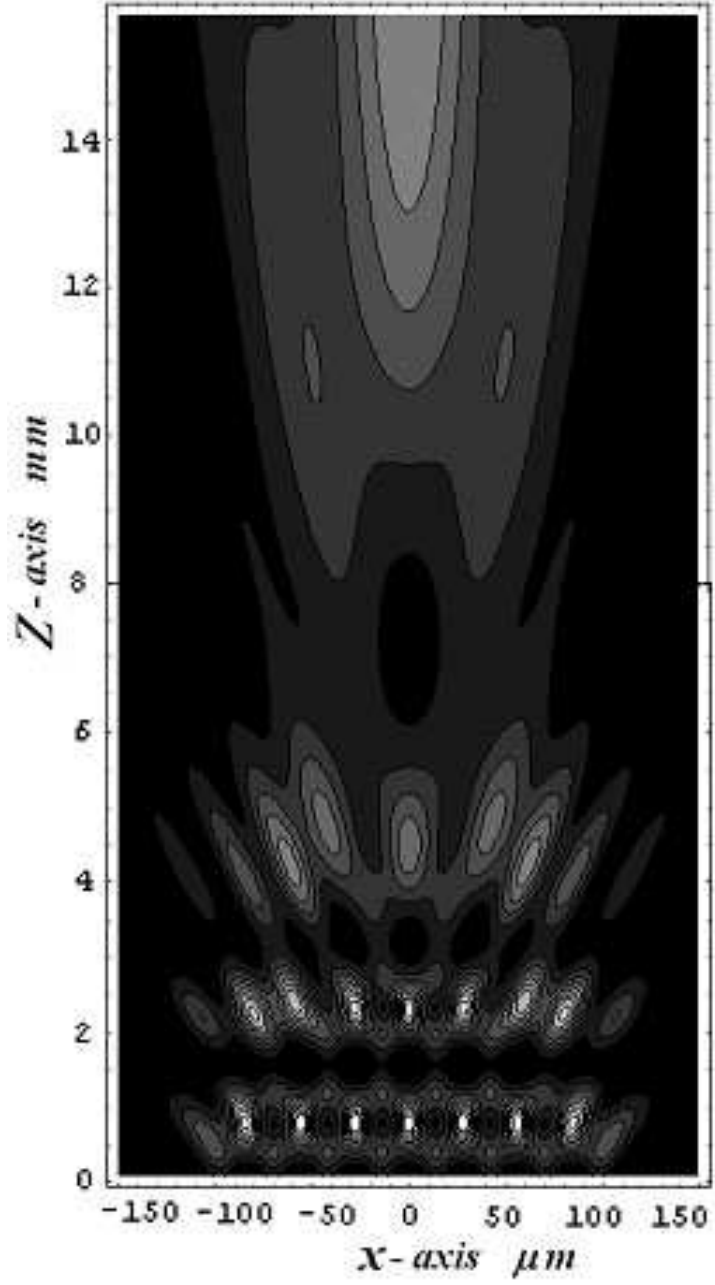}}

\caption{Diffractive self-imaging of two-dimensional lattice of 
8x8 Gaussian beams with period $p=28{\mu}m$ . The longitudinal
cross-section at $y=0$ plane of intensity distribution $I(x,y,z)$ presented. The lattice 
is self-reproduced 
 at Talbot distance, spatial period division occurs at quarter Talbot 
distances and central lobe forms outside the Rayleigh range.}
\label{fig.2}
\end{figure}

\begin{figure}

\center{\includegraphics[width=.8\linewidth] {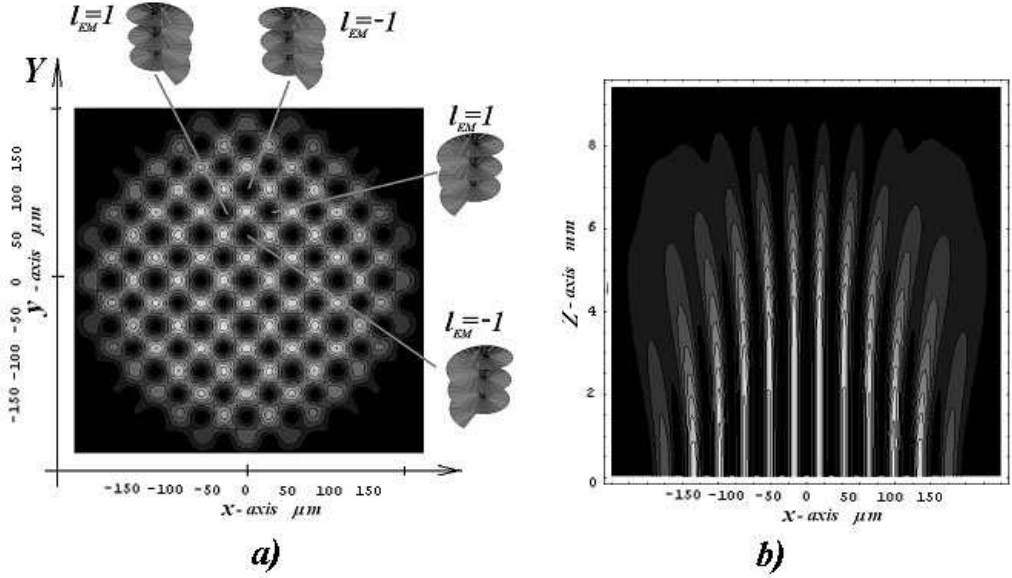}}

\caption{a) Intensity distribution in transverse plane of artificial Laquerre-Gaussian vortex array. 
Vortices(dark holes) are at nodes of rectangular $chessboard-like$ 
grid. Letters $ \ell_{EM}$ denote vortices with alternating topological charges,
changing sign from one site of lattice to another. b)Longitudinal section of vortex array in the near field in $x$-$z$ plane at $y=0$ section. 
The vortex lines are parallel, the topological charges $\ell_{EM}$ are flipping 
from one vortex line to another. The wavelength $\lambda = 1 \mu m$, period $p$ of lattice
both in $x$ and $y$ directions is $30 \mu m$ }

\label{fig.3}
\end{figure}

\begin{figure}
\center{\includegraphics[width=.6\linewidth] {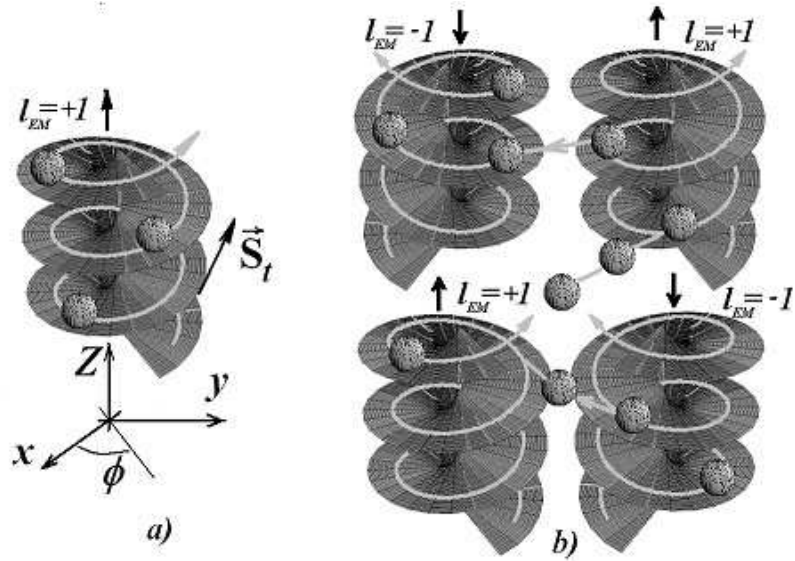}}

\caption{ a)Helicoidal phase surface of Laquerre-Gaussian beam. 
Atomic dipole having "red" detuning moves along phase gradient. Trajectory is located at 
maximum of intensity. $  \vec{S_t}  $  is the component of Pointing vector tangent to helix.
The major component of $  \vec{S_z}  $ is directed parallel to beam propagation, i.e. along 
$Z$-axis. b)Helicoidal phase surface of Laquerre-Gaussian beam array. 
The elementary cell consists of four vortices. 
The ajaicent vortices have arternating topological charges $\ell_{EM}$
and alternating  angular momenta. In classcal picture the 
atomic dipole moves along phase gradient of a given vortex 
next it jumps to another one.}
\label{fig.4}
\end{figure}

\begin{figure}

\center{\includegraphics[width=.8\linewidth] {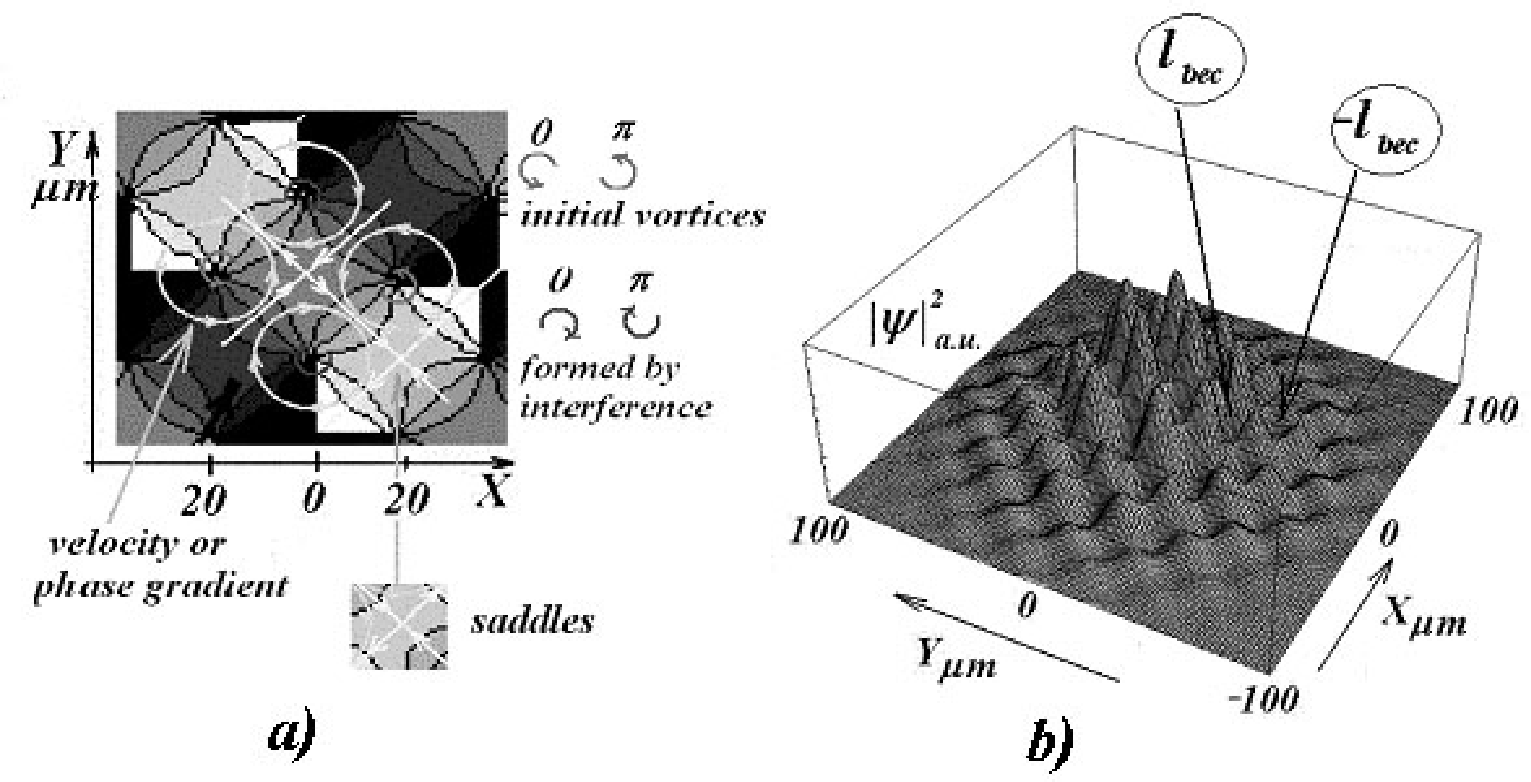}}

\caption{a)The argument of macroscopic wavefunction $\Psi$ and corresponding 
field of velocities obtained via Madelung transform . The superfluid vortices form the lattice 
with alternating topogical charges $\ell_{BEC}=\pm 1$. The elementary cell 
consists of four BEC vortices whose locations are identical to the 
vortices of trapping EM-field. Horizontal pair has the same charges  $\ell_{BEC}$
which are $\pi$-shifted with respect to each other, the vertical pair 
have  $\pi$-shifted $-\ell_{BEC}$ charges.
The altenating charges $\ell_{BEC}$ make the field of velocities continuous.b)
 The distribution of the 
square modulus of macroscopic wavefunction $\Psi$ obtained via numerical modeling.
The location of adjaicent counter-rotating vortices $\ell_{BEC}$ shown.}
\label{fig.5}
\end{figure}

The additional parabolic well with frequency ${\omega{_{\bot}}}$ 
is introduced in (\ref{TRAPOP})  to get analytical solution for one trapping vortex. 
In order to avoid the interference between different trapping beams
the usage of different carrier frequencies  is recommended for longitudinal parabolic well 
${\frac { m{\:} {\omega{_z}}^2 {z^2}}{2 }}$,  vortex array beam ${E}(\vec{r} )$
and parabolic subtrap in (\ref{TRAPOP}).The characteristic scales of potential in longitudinal direction and transverse direction are choosen 
to form "pancake trap" : ${\omega{_z}}>>\omega_{\bot}$ (fig. \ref{fig.1}). The opposite case of elongated trap 
with ${\omega{_z}}<<\omega_{\bot}$  and "solitonic" longitudinal potential $V_z \approx {sech}^2(z)$ was
considered earlier using analogous procedure \cite{Manko:2004}.

The longitudinal part of wavefunction ${\Psi_{||}}(z,t )$ is obtained as a ground state of 1D harmonic oscillator:

\begin{equation}
\label{osc1d}
\ {{ {i}} \hbar}{\:}{\frac {\partial {\Psi_{||}}}{\partial t}} =
 -{\frac {\hbar^2}{2 m}} {\frac {\partial^2 {\Psi_{||}}}{\partial z^2}}+ {\frac { m{\:} {\omega{_z}}^2 {z^2}}{2 }}{\Psi_{||}}
\end{equation}

\begin{equation}
\label{ground1D}  
\ {\Psi_{||}}={{ ( \frac {m \omega_z }{ \pi \hbar})^{1/4} }}{\:}exp{\:}[-m \omega_z z^2/ (2 \hbar)- {{ {i}}{\:}\omega_{z}  {\:}t}]
\end{equation}

The transverse  part of wavefunction ${\Psi_{\bot}  (\vec{r}_{\bot},t ) }$
is to be obtained by solving "transverse" GPE:

\begin{eqnarray}
\label{GPET}
\  {{ {i}} \hbar}{\:}{\frac {\partial {\Psi_{\bot}}}{\partial t}} =
 -{\frac {\hbar^2}{2 m}} \Delta_{\bot} {\Psi_{\bot}}+V_{\bot}(\vec{r_{\bot}}){\Psi}_{\bot}
&+&\nonumber \\
{\frac{4{\pi}{\hbar}{\:}^2{\:} {a( {\vec{B}})}}{m}}
 {\Psi_{\bot}}|{\Psi_{\bot}} |{\:}^2 [ \int^\infty_{-\infty}|{\Psi_{||}}(z,t ) |{\:}^4 d z]/
[  \int^\infty_{-\infty}|{\Psi_{||}}(z,t ) |{\:}^2 d z],
\end{eqnarray}

where

\begin{equation}
\label{potential}
\ {V_{\bot}}= {\frac { m{\:} {\omega{_{\bot}}}^2  | \vec{r_{\bot}}  |{\:}^2 } {2 }}-
Re[{\alpha(\omega)}] {\:}
 {|{E}(\vec{r_{\bot}} ) |{\:}^2}
\end{equation}
Because of normalization 

\begin{equation}
\label{normapsi}  
\int^\infty_{-\infty}|{\Psi_{||}}(z,t ) |{\:}^4 d z= 1/2 {\:}{\:}{\:}{\:}and {\:}{\:}\int^\infty_{-\infty}|{\Psi_{||}}(z,t ) |{\:}^2 d z= 1
\end{equation}

The following 2D GPE results from separation of variables for "pancake" trap:

\begin{equation}
\label{GP2D}
\  {{ {i}} \hbar}{\:}{\frac {\partial {\Psi_{\bot}}}{\partial t}} =
 -{\frac {\hbar^2}{2 m}} \Delta_{\bot} {\Psi_{\bot}}+V_{\bot}(\vec{r_{\bot}}){\Psi}_{\bot}+
{\frac{2{\pi}{\hbar}{\:}^2{\:}{a({\vec{B}})}}{m}}{\Psi_{\bot}}|{\Psi_{\bot}} |{\:}^2
\end{equation}

The scattering length $  {a}$ as a function of magnetic field is:

\begin{equation}
\label{Feshbach}
  {a(\mathbf {\vec{B}})}=   { {a}_{bg}}{\:} ({1+\frac{{  {\Delta_{  {B}}}}}{  {B-B_{R}}}})
\end{equation}
where $  {\Delta_{  {B}}}$ is a width of Feshbach resonance, $  {B_{R}}$ - resonant magnetic field,
$   {a_{{\:}bg}}$-background scattering length\cite{Pitaevskii:2006}.

Formally the separation of variables is applicable each time when Hamiltonian is
a sum of components depending on different  groups of variables, 
but this method have additional physical meaning 
for asymmetric potentials. The examples are elongated in $z$-direction trap \cite{Manko:2004} and  "pancake" 2D trap (fig. \ref{fig.1})
as in current case. The dynamics of $\Psi$ in 2D traps was considered in large amount 
of papers, including the geometries of periodic potentials, Bessel lattices etc. In current case 
the SFV lattice under consideration has some features, qualitatively described above 
in discussion of classical motion of dipole around phase singularity.

The continuous transfer of angular momentum from optical vortex to 
BEC with wavefunction $\Psi$ might induce the vortices in initially nonrotating BEC. 
Because the direction of rotation of classical particle is determined by 
distribution of TC of optical vorices $\ell_{EM}$ of trapping beam the distribution of TC's 
in quantum  superfluid  lattice ${ \ell_{BEC}}$ will be set  identical to those of trapping field.

The possible solution of  equation  (\ref{GP2D}) presumes the identical spatial distributions for
fields $ {\Psi_{\bot}}(\vec{r_{\bot}},t )$ and ${E}(\vec{r_{\bot}} )$. It means that $\Psi$ is also a 
sum of a LG functions with alternating topological charges${\:}{\:} \ell_{BEC}$ (see eq.(\ref{psi_vortex_array})).
Unfortunately the effective diameter of the core $b$ is much larger than effective size of LG beam 
bottleneck (4-10 $\mu m$). But this discrepancy does affect the basic features of solution because
of logarythmic dependence of vortex energy on vortex dimensions \cite{Feynman:1972}.

The correlation $K$ of two complex spatially inhomogeneous fields $ {\Psi_{\bot}}(\vec{r_{\bot}},t )$ and ${E}(\vec{r_{\bot}} )$ 
is expected to be unit:

\begin{equation}
\label{correlation}  
\ { K} {\:}={\frac{ |{\int \Psi_{\bot}{E}^{*}(\vec{r_{\bot}} ){d^{{\:}2}{\vec{r_{\bot}}}}}|^2}{
{[{\int |\Psi_{\bot}|^2{d^{{\:}2}{\vec{r_{\bot}}}}}}] [{\int  |{E}(\vec{r_{\bot}} )|^2{d^{{\:}2}{\vec{r_{\bot}}}}}]}
 }=1
\end{equation}

The key point is in adjusting the 
parameters in  $ {\Psi_{\bot}}(\vec{r_{\bot}},t )$ and ${E}(\vec{r_{\bot}} )$ in such a way that two last terms in 
(\ref{GPET}) would cancel each other.This might happen when  following condition is imposed upon the coefficients: 
\begin{equation}
\label{normapsi1}  
\ Re[{\alpha(\omega)}] {\:}
|{E_0} |{\:}^2 {\:}={\frac{2{\pi}{\hbar}{\:}^2 {\:}   {a(  {\vec{B}})}}{m}}
\end{equation}

The mutual compensation of these two terms could be achieved via tuning of magnetic field $ {\vec{B}}$ 
near Feshbach resonance.

Consider first the case of single vortex trap collocated with a single parabolic subtrap:
 \begin{equation}
\label{single_vortex_trap}
\ {V_{\bot}}= {\frac { m{\:} {\omega{_{\bot}}}^2   r {\:}^2 } {2 }}-Re[{\alpha(\omega)}] {\:}|{E_0} |{\:}^2 {\:} r^2  {\:}exp{\:}[ {-r^2}/d^{{\:}2}]
\end{equation}
The following  exact solution for 2D harmonic oscillator is known :
\begin{equation}
\label{LG_single_wavefunction}
\ {\Psi_{\bot}}={{\sqrt\frac {2}{ \pi} }}{{[\frac {m \omega_{\bot} }{ \hbar} ]^{3/2} }}{\:}r{\:}exp{\:}[-{\frac 
{m {\:}{\omega_{\bot}}^2  r {\:}^2}{2 \hbar}}+{\:}{  {i}}  {\:}\phi  {\:}{ \ell_{BEC}} {\:}-{{  {i}}2{\:}\omega_{\bot}  {\:}t}]
\end{equation}
which is similar to variational solution for "transverse"  wavefunction (\ref{LG}) \cite{Abraham:2002}.
Note in our case the exact wavefunction of the transversal GPE found, rather than variational one (\ref{LG}).
The stability analysis will be published elsewhere.The angular momentum per particle is given by:

\begin{equation}
\label{2Daom}
\ {<{ \hat{ \ell}}>} = \int \int  {\Psi^{*}_{\bot}} {{{( {-}i}} \hbar)}{\:}{\frac {\partial {\Psi_{\bot}}}{\partial \phi}} d^{{\:}2} {\vec{r_{\bot}}}=\hbar
 \end{equation}
where $\phi$ is azymuthal angle (fig. \ref{fig.4a}).
Again SFV carries angular momentum $\hbar$ per particle and kinetic energy  of the whole vortex line
${ E_{kin}= {\rho}{\:} \pi {\hbar}^2{\:} {\chi}{\:}  ln(\tilde{b} / \tilde{a})/m}$ \cite{Feynman:1972}.Consider now 
the trapping of BEC by of phase-locked Gaussian-Laquerre beams
 placed at the nodes $i,j$ of rectangular grid of period $p$ ( see Eq. \ref{hold_vort}).

 Let us assume for simplicity that optical wavelength $\lambda$
is equal to De-Broiglie wavelength $\lambda _{db}$: 

\begin{equation}
\label{DeBroiglie}
\  \lambda = \lambda _{db} =
{\frac {\hbar}{\sqrt{2 m k_B T}}}
\end{equation}

The corresponding BEC temperature for ${^{85}}Rb$ atoms is:

\begin{equation}
\label{BEC_temp}
\  T = {1.7}^{-7} K
\end{equation}

After imposing compensation condition (\ref{normapsi}) the residual part of transverse GPE (\ref{GP2D}) corresponds to free-space evolution:

\begin{equation}
\label{2Dfreespace}
\  {{  {i}} \hbar}{\:}{\frac {\partial {\Psi_{\bot}}}{\partial t}} =
-{\frac {\hbar^2}{2 m}} \Delta_{\bot} {\Psi_{\bot}}+ {\frac { m{\:} {\omega{_{\bot}}}^2   r {\:}^2 } {2 }}{\Psi_{\bot}}
\end{equation}

The free space propagation (fig. \ref{fig.1}) equation for EM-field will be of similar form:

\begin{equation}
\label{2Dfreespace_EM}
\  {\:}{\frac {\partial {E}}{\partial z}} =
-{\frac {   {i}}{2 k}} \Delta_{\bot} {E}+ {\frac { k {\:}  r {\:}^2 } {2 f_{cavity}}} {E}
\end{equation}

where $f_{cavity}$ is effective focal length of the laser cavity  (fig. \ref{fig.1}) induced by thermal lense
collocated with inhomogeneity of optical pumping.

Because of linearity of equation and superposition principle the solution of ( \ref{2Dfreespace_EM} )  will be the sum
of free-space modes including  zero-order Gaussian functions, Gaussian-Hermit or Laguerre-Gaussian modes
located at sites $\vec{r}_{jx,jy}$ \cite{Okulov:1990} separated by period $p$.
The axes of such optical array are parallel to $X$,$Y$.
In order to solve numerically eq. (\ref{GP2D}) the split-step FFT method \cite{Okulov:1993}
was used. The mesh size in $X,Y$ plane was $512 \cdot 512$ points,
the "guard bands ratio" \cite{Siegman:1975} was chosen equal to $8$. So the main part of the field
$ \Psi $ was located inside central part of a mesh which had  $64 \cdot 64$ size - the "image area". The tolerance 
of the energy spillover was kept within $\epsilon_1 = 0.0001$.The windowing in 
wavenumber space after FFT at each timestep was performed by usage of "Fermi-Dirac" smoothed
step function \cite{Okulov:1993}. The dissipation inherent to split-step FFT method have led to 
decrease of total "amount of particles" $\int \int {|{\Psi}|^2 }dx dy $ within "image area" at a speed 
of $10^{-3}$ per time step.
The special initial conditions of "preselected" SFV array in the form (\ref{psi_vortex_array}) superimposed 
upon homogeneous background gave the
spatial distribution of "transverse" wavefunction $\Psi_{\bot}$ well correlated ($K \approx 0.7 $) with 
the OVA array distribution - the array of phase-locked Gaussian-Laquerre wavepackets of the first order : 
\begin{eqnarray}
\label{LG_array}
\ {\Psi_{\bot}}={{\frac {2}{ \pi} }}{{[\frac {m \omega_{\bot} }{ \hbar} ]^{3/2} }}{\:}\sum_{jx,jy}(|{\vec{r}-\vec{r}_{jx,jy}} | ){\:}
{ exp{{\:}(- {\:} | \vec{r} |{\:}^2/D^2})}{\:} 
exp{\:}[-{\frac 
{m {\:}{\omega_{\bot}}^2 | {\vec{r}-\vec{r}_{jx,jy}} |{\:}^2}{2 \hbar}}
&+&\nonumber \\ {\:}{  {i}}{\:}  \ell_{BEC} Arg( \vec{r}-\vec{r}_{jx,jy} ){\:}+ {\:}{  {i}}{\:} \pi (jx+jy) - {{  {i}}2{\:}\omega_{\bot}  {\:}t}]
\end{eqnarray}

Thus the numerical solution for ${\Psi_{\bot}}$ of GPE proved to be very close to the linear combination of 
Gaussian-Laquerre functions. Nevertheless this solution takes into account the interference between 
wavefunctions of "subvortices", because the "doughnut" radius 
is set to be a bit more than distance between lattice nodes. The arising intereference pattern is well correlated 
with the interference pattern produced by Laguerre-Gaussian 
OVA $|{{E}(\vec{r_{\bot}} )} |{\:}^2 {\:} $ (fig. \ref{fig.3}) with the same geometrical parameters
and wavelength $\lambda$. This solution proved to be stable with respect to small harmonic perturbations.

Each SFV carries angular momentum $\hbar$ per particle and rotational kinetic energy 
${ E_{kin}= {\rho}{\:} \pi {\hbar}^2{\:} {\chi}{\:} ln(\tilde{b} / \tilde{a})/m}$, $\tilde{b}$ - diameter of vortex core, 
$\tilde{a}$ - interatomic distance, ${\chi}$ - length of vortex which is roughly equal 
to thickness of  "pancake"  \cite{Feynman:1972}. The energy associated with superfluid vortices is of 
order $10^{-(19 \div 20)}J$ or $0.1 \div 1 {\:}eV$ at density of dilute ${^{85}}Rb$ gas ${{\rho}{\:} \approx 10^{16 \div 18}} cm^{-3}$.
In contrast to superfluid in rotating  bucket where angular momenta of vortices are 
co-directed \cite{Ketterle:2002} , the BEC vortices trapped by optical vortex array are counter-directed from site to site. 
Consequently in the net sum of angular momenta each vortice of positive topological charge is compensated by 
the term with negative charge and total angular momentum tends to zero.
Nevertheless the mutual
subtraction of angular momenta(vectorial) of adjaicent vortices in the net sum does not mean the mutual 
subtraction of rotational energies, which are the positive scalars. The ground state carry  substantial 
amount of rotational kinetic energy of condensate which contains $N$ particles of mass $m$ per unite volume, 
namely ${ E_{rot}= {N_{vortices}} {\:}{\rho}{\:} \pi{\:} {\hbar}^2 {\:} {\chi}{\:} ln(\tilde{b} / \tilde{a})/m}$  . 

\section{\label{sec:level1}Conclusion}

		The optical vortex arrays emitted by solid-state microchip laser were analysed from the point of view 
of application to optical dipole traps. Firstly the numerical modeling of thin slice microchip
Fabry-Perot solid-state laser resonator gave transverse field distributions(fig. \ref{fig.3}) well correlated with 
experimentally observed previously. The longitudinal intensity distribution consists of periodically
spaced array of parallel hollow tubes which slowly diverge while propagating along $z$-axis(fig. \ref{fig.3}).  
The array of phase-locked Gaussian-Laquerre beams equispaced at the nodes of rectangular lattice proved to be 
a reasonable approximation for experimental and numerical results as well .   

For macroscopic wavefunction $\Psi$ of BEC trapped in such complex optical field the analytical  
solutions of Gross-Pitaevsky
equation were found. These solutions are based upon separation of variables and 
mutual compensation of vortex component of 
external trapping field via nonlinear term of GPE. The obtained wavefunctions have perfect correlation with 
trapping field, including distribution of topological charges, which form "antiferromagnetic-like" lattice.
Within framework of this particular model the "antiferromagnetic" lattice of BEC vortices carriers total 
angular momentum close to zero while net rotational kinetic energy of SFV lattice tends to be equal to the sum 
of rotational energies of vortices. Geometrically such BEC - cloud looks like "pancake" perpendicular to
$z$-axis.

The field of classical velocities, i.e. the field of phase gradients, obtained via Madelung transform,
forms "labyrinth" structure. It means that  trapped atoms move in the 
"pancake" plane, i.e. $x-y$ plane.The rotation of atom around some vortex
core is accelerated by EM-torque. The radius of rotation is gradually increased. When particle
reaches the separatrice it comes to another 
vortex area. The trajectory of particle 
in transverse plane ($x-y$ plane) is Mobius-like: 
because the number of optical vortices is finite,
in classical picture the particle will return to initial 
vortex eventually after roaming for some time 
inside trapping "labyrinth".

In quantum picture represented via analytic solution of GPE for one trapping vortex and numerical 
solution for optical vortex array the 
coherent macroscopic wavefunction extends all over OV trapping array with transverse spatial 
dimension of several hundred microns.  The complex field of velocities,  rotational 
energy and high degree of correlation of SFV wavefunction with OV trapping field promise 
more resistance to decoherence. 

The qualitative analytic solution supports the basic feature related to 
trapping of macroscopic particles and to BEC trapping: the transfer of OAM from optical field 
to superfluid. The proposed OVA trap  might be interesting from the point of view of studies 
of quantum-classical correspondence.

The mechanism of imposing the topological charges to BEC vortices by means of manipulating 
the vorticity of trapping optical array could result in demonstration of macroscopic quantum
interference phenomena. Evidently there are 4 possible topologically equivalent combinations 
of parameters of solution (\ref{LG_array}) of OV charges ($ \ell_{EM}=\pm {\:}1$ ) and their relative phases ($\pm {\:}\pi$ ).
Thus there exist 4 wavefunctions $\Psi$ having an indentical probabability $|\Psi|^2$ distribution and 
different orientation of SFL vortices
with respect to physical axes $x,y$ of trapping setup, characterised by $ \ell_{BEC}$ and their relative phases.
The transformation of the one $\Psi$ into another one which have different phase structure is equivalent 
to $90^{\circ}$ rotation around $z$ - axis.

\newpage

\section*{List of Figure Captions}

Fig.1{Conceptual view of the near field optical trap. Upper plot:
 Transverse (in XOY plane) distribution of intensity in the near field of
solid-state microchip laser {\cite{Okulov:2003}}. 
Middle: The longitudinal scale extends to 6 Talbot lengths. 
Z - axis is directed along optical axis of microchip laser 
resonator(bottom) . Additional tightly confined parabolic well
keeping BEC cloud localized in $z$ - direction is depicted 
via potential $V(z) \approx (z-z^{\prime})^2$. Such potential is assumed to be 
superimposed by the other microchip laser beam with cylindrical focusing lens at slightly different
wavelength 
of radiation from the range ${0.98 \div 2.79 {\mu} m}$, in order to avoid interference.
 Fresnel number has value of $N_f \approx {N_{vortices}} ^2    =  64$.}

Fig.2   {Diffractive self-imaging of two-dimensional lattice of 
8x8 Gaussian beams with period $p=28{\mu}m$ . The longitudinal
cross-section at $y=0$ plane of intensity distribution $I(x,y,z)$ presented. The lattice 
is self-reproduced 
 at Talbot distance, spatial period division occurs at quarter Talbot 
distances and central lobe forms outside the Rayleigh range.}

Fig.3{\:}{a) Intensity distribution in transverse plane of artificial Laquerre-Gaussian vortex array. 
Vortices(dark holes) are at nodes of rectangular $chessboard-like$ 
grid. Letters $ \ell_{EM}$ denote vortices with alternating topological charges,
changing sign from one site of lattice to another. b)Longitudinal section of vortex array in the near field in $x$-$z$ plane at $y=0$ section. 
The vortex lines are parallel, the topological charges $\ell_{EM}$ are flipping 
from one vortex line to another. The wavelength $\lambda = 1 \mu m$, period $p$ of lattice
both in $x$ and $y$ directions is $30 \mu m$ }

Fig.4{\:}{ a)Helicoidal phase surface of Laquerre-Gaussian beam. 
Atomic dipole having "red" detuning moves along phase gradient. Trajectory is located at 
maximum of intensity. $  \vec{S_t}  $  is the component of Pointing vector tangent to helix.
The major component of $  \vec{S_z}  $ is directed parallel to beam propagation, i.e. along 
$Z$-axis. b)Helicoidal phase surface of Laquerre-Gaussian beam array. 
The elementary cell consists of four vortices. 
The ajaicent vortices have arternating topological charges $\ell_{EM}$
and alternating  angular momenta. In classcal picture the 
atomic dipole moves along phase gradient of a given vortex 
next it jumps to another one.}

Fig.5{\:}{a)The argument of macroscopic wavefunction $\Psi$ and corresponding 
field of velocities obtained via Madelung transform . The superfluid vortices form the lattice 
with alternating topogical charges $\ell_{BEC}=\pm 1$. The elementary cell 
consists of four BEC vortices whose locations are identical to the 
vortices of trapping EM-field. Horizontal pair has the same charges  $\ell_{BEC}$
which are $\pi$-shifted with respect to each other, the vertical pair 
have  $\pi$-shifted $-\ell_{BEC}$ charges.
The altenating charges $\ell_{BEC}$ make the field of velocities continuous.b)
 The distribution of the 
square modulus of macroscopic wavefunction $\Psi$ obtained via numerical modeling.
The location of adjaicent counter-rotating vortices $\ell_{BEC}$ shown.}

\newpage

\begin{thebibliography}{99}

{\small

\bibitem{Ovchinnikov:2000}{ R.Grimm, M.Weidemuller and Yu.B.Ovchinnikov, }
\textit{Adv.At.Mol.Opt.Phys.} {\bf 42}, 95 (2000).

\bibitem{Pitaevskii:2006}
{L. Pitaevskii and S. Stringari,} \textit {"Bose-Einstein Condensation"},Clarendon Press, Oxford, {2003}.

{F. Dalfovo, S.Giorgini, S.Stringari, L.P.Pitaevskii, }{Rev.Mod.Phys.} {\bf 71}, 463 (1999).

\bibitem{Letokhov:1988}
{Balykin, V.I., Letokhov, V.S., Ovchinnikov, Yu.B., and Sidorov A.I,} 
{Phys. Rev. Lett. } {\bf 60},  2137 (1988).

\bibitem{Wright:2001} E. M. Wright, J. Arlt and K. Dholakia,{Phys.Rev. A} {\bf 63}, 013608 (2001).

\bibitem{Okulov:2003}
{A.Yu.Okulov,} \textit {"3D-configuration of the vortex lattices in microchip laser cavity"}, 

 QCMC-2004, AIP Conference Proceedings, {\bf 734}, p.366 (2004).

{A.Yu.Okulov,}Bulletin Lebedev Physical Institute, {\bf 9}, p.3, Sept. (2003).

\bibitem{Ovchinnikov:2006}
{Yu.B.Ovchinnikov,}Phys.Rev, {\bf 73A}, 033404 (2006).

\bibitem{Ertmer:2002}
{R.Dumke, M.Volk, T. Muther, F.B.J.Buchkremer, G. Birkl and W.Ertmer,}
{Phys. Rev. Lett. } {\bf 89}, 097903 (2002).


\bibitem{Allen:1994}
{M. Babiker, W. L. Power, and L. Allen,}{Phys. Rev. Lett. } {\bf 73}, 1239 (1994).

\bibitem{Padgett:2003}
{V.Garces-Chavez,D. McGloin, M. J. Padgett, W. Dultz,
 H. Schmitzer, and K. Dholakia,}
{Phys. Rev. Lett. } {\bf 91}, 093602 (2003).

\bibitem{Dunlop} M.J.Friese,J.Euger,H.Rubinstein-Dunlop, {Phys.Rev. A} {\bf 54}, 1543 (1996).

{N.R. Heckenberg, M.E.J. Friese, T.A. Nieminen and H. Rubinsztein-Dunlop,}
{"Mechanical Effects of Optical Vortices"}.

pp 75-105 ,in M. Vasnetsov (ed), \textit {Optical Vortices},
Nova Science Publishers (1999).

\bibitem{Abraham:2002}
{J. Tempere and J. T. Devreese, E. R. I. Abraham.,} 
Phys.Rev., {\bf 64A}, 023603 (2002).


\bibitem{Okulov:1990}
{A.Yu.Okulov,} {JOSA} {\bf B7}, p.1045, (1990).

\bibitem{Courtial:2006}
{J.Courtial,R.Zambrini,M.R.Dennis,M.Vasnetsov,} Optics Express, {\bf 14}, p.938 (2006).

\bibitem{Chen:2001}{Y.F.Chen, Y.P.Lan,} 
Phys.Rev., {\bf 64A}, 063807 (2001).

{Y. F. Chen,Y. P. Lan,}
Phys.Rev., {\bf 65}, 013802 (2001).

\bibitem{Manko:2004}
{R.Fidele,P.K.Shukla,S.De.Nicola,M.A.Manko,V.I.Manko,F.S.Cataliotti,} 
JETP Lett., {\bf 80},8, p.609-613 (2004).

\bibitem{Ketterle:2002}
{J.R.Abo-Shaerr, C.Raman, J.M.Vogels and W.Ketterle,}
{Phys. Rev. Lett. } {\bf 88}, 070409 (2002).
{I.Danaila,{Phys.Rev. A} {\bf 72}, 013605  (2005).}

\bibitem{Feynman:1972}
{R.P.Feynman,} "Statistical mechanics", Ch.11, Reading, Massachusetts (1972).

\bibitem{Kitaev:2001}
{A.Yu.Kitaev,} LANL e-print quant-ph/ 9707021, http://arxiv.org (1997).

\bibitem{Suchkov:1965} A.F.Suchkov, {JETP} {\bf 22}, 1026 (1965).
\bibitem{Siegman:1986} A.E.Siegman, {\it "Lasers". University Science Books, Mill Valley, California}(1986).

\bibitem{Staliunas:1993} K.Staliunas,{Phys.Rev. A} {\bf 48}, 1573 (1993).
K.Staliunas, C.O.Weiss, {JOSA B} {\bf 12}, 1142  (1995).

\bibitem{Hollinger:1990} F. Hollinger, Chr. Jung, and H. Weber,{JOSA B} {\bf 7}, 1013 (1990).

\bibitem{Okulov:1993}
{A.Yu.Okulov,} Opt.Comm.{\bf 99}, p.350-354 (1993).
J.Mod.Opt.{\bf 38},N.10,p.1887(1991).

\bibitem{Okulov:1994}
{A.Yu.Okulov,}Optics and Spectroscopy, {\bf 77}, N6,p.985 (1994).

\bibitem{Dholakia:2002}
{V. Garces-Chavez, K. Volke-Sepulveda, S. Chavez-Cerda, W. Sibbett and K. Dholakia,} 
{Phys. Rev.A } {\bf 66}, 063402 (2002).

\bibitem{Siegman:1975}
{E.A.Sziklas,A.E.Siegman,}  Appl.Opt. {\bf 14}, 1874 (1975).

}
\end{thebibliography}
\end{document}